\documentclass[preprint,tightenlines,eqsecnum,floats,aps,amsmath,amssymb,nofootinbib,prd,showpacs]{revtex4}

\usepackage{amssymb}
\usepackage{stmaryrd}
\usepackage{amsmath}
\usepackage{amsfonts}
\usepackage{mathrsfs}
\usepackage{CJK}
\usepackage{amsmath,amssymb,amsfonts}
\usepackage{graphicx}
\usepackage{subfigure}
\usepackage{color} 

\def\be{\begin{equation}}
	\def\ee{\end{equation}}
\def\ba{\begin{eqnarray}}
	\def\ea{\end{eqnarray}}
\def\nn{\nonumber}

\begin{document}
	
	\title{Topological classes of BTZ black holes}
	
	\author{Yongbin Du}
	\affiliation{Department of Physics, South China University of Technology, Guangzhou 510641, China}

	\author{ Xiangdong Zhang\footnote{Corresponding author. scxdzhang@scut.edu.cn}}
	\affiliation{Department of Physics, South China University of Technology, Guangzhou 510641, China}

	\date{\today}
	
	
	\begin{abstract}
In the recent paper [Phys. Rev. Lett. 129, 191101 (2022)], the black holes were viewed as topological thermodynamic defects by using the generalized off-shell free energy. Their work indicates that all black hole solutions in the pure Einstein-Maxwell gravity theory could be classified into three different topological classes for four and higher spacetime dimensions. In this paper, we investigate the topological number of BTZ black holes with different charges $(Q)$ and rotational $(J)$ parameters. By using generalized free energy and Duan's $\phi$-mapping topological current theory, we interestingly found only two topological classes for BTZ spacetime. Particularly, for $Q=J=0$ BTZ black hole, there has only one zero point and therefore the total topological number is 1. While for rotating or charged cases, there are always two zero points and the global topological number is zero.
	\end{abstract}
	\maketitle

	\section{INTRODUCTION}
	
	As one of the most fascinating objects
	in nature, the black hole has receiving increasingly attention both in theoretical and observational physics. On one hand, singularity theorem \cite{Penrose} reveals the intrinsic limitation of general relativity. On the other side, when taking quantum effect into consideration, people found that black hole is not black at all \cite{Hawking} and surprisingly sharing rich thermodynamical properties \cite{thermo1,thermo2,thermo3}. Many issues, such as information paradox \cite{paradox}, arise due to these remarkable discoveries.
	
	Recently, by introducing Duan's
	topological current $\phi$-mapping theory\cite{Duan}, Wei et. al. \cite{Liu22,Liu21} originally consider black hole as topological thermodynamic defects, and classify different solutions with their global topological charge. They divided black hole solutions into three different topological classes according to
	their different topological numbers. This method gives us an insight into black hole solutions and implies more inherent properties of this extraordinary thermodynamic system.
	
	Inspired by this results, many works have been done to investigate the topological charge of different types of black hole. Since \cite{Liu22} only foucus on the static solutions, in \cite{Kerr}, stationary black holes are explored. The topological number of Kerr and Kerr-Newman black hole are calculated. Furthermore, they calculate the singly-rotating black hole in higher dimensions \cite{Kerr}. Apart from the researches on black hole in general relativity, some researchers are also consider the black hole in modified gravity theories such as Gauss-Bonnet gravity \cite{GB} or Lovelock gravity \cite{Lovelock} and so on \cite{Bham1,Bham2,Bham3}. The topological number are quite different in these cases, as provide us a new perspective to consider the difference between modified gravity theory and general relativity.
	
	An interesting feature of these works is that whether we consider black hole solutions in general relativity or modified gravity or even high-dimensional extension, they can always be classified into three topological classes \cite{Kerr,GB,Lovelock}. Moreover, dimension may play an improtant role in the amount of zero points and topological number \cite{Kerr,GB}. Hence extending the approach to lower dimensional case could provide us with a test to the seemingly universal phenomenon or provide a counterexample which become very relevant for the future study on this issue. To this aim, in this paper, we follow this newly-hewed path to study the topological number of the Banados-Teitelboim-Zanelli (BTZ) black hole \cite{BTZ} and trying to provide counterexample to the above observation.
	
	The BTZ black hole is the black hole solution of 3-dimensional general relativity plus a negative cosmological constant and plays an important role in quantum theory of gravity. As a low-dimensional toy model, BTZ black hole have many interesting features. Locally, it has constant negative curvature and consequently no curvature singularity \cite{nosingularity}.  But it possess a horizon, and the Bekenstein's area law could also be applied in it. Thus it is a genuine black hole and the simply form may benefit quantum gravity or holographic priciple a lot. Many paper focus on this 3-dimensional black hole solution to find some unexpected result of common physical phenomenon, such as Penrose process \cite{zhang1,zhang2}.  In this article we shall concentrate on thermodynamics of BTZ black hole and then calculate the topological charge when the angular momentum and electric charge are taken different values. We would also like to give a simple comparison with the results of black hole in four dimensions \cite{Kerr,Liu22,Liu21}.

This paper is organized as follows: In section 2, we briefly review some useful results of thermodynamics of BTZ black hole. Then we use the generalized free energy to establish a parameter space in section 3, and find the zero points with their topological charge. In section 4, we summarize our result, compare it with higher dimensional cases and make some comment on this new approach.

	\section{Thermodynamics of BTZ black hole}
	In this section, we first give a brief review of thermodynamics of BTZ black hole. We use the generalized off-shell free energy \cite{York} so that we could classify different black hole solutions. This means black holes with the same energy $($and electric charge or angular momentum, if any$)$ can be in different temperature. The metric line element for the BTZ black hole reads \cite{BTZ}
	
	\begin{equation}
		d s^2=-N^2 d t^2+N^{-2} d r^2+r^2\left(N^\phi d t+d \phi\right)^2,
	\end{equation}
	where the squared lapse $N^2(r)$ and the angular shift $N^\phi(r)$ are respectively given by
	\begin{eqnarray}
		N^2(r)&=&-M+\frac{r^2}{L^2}+\frac{J^2}{4 r^2},\\
		N^\phi(r)&=&-\frac{J}{2 r^2}
	\end{eqnarray}
	with $-\infty<t<\infty$, $0<r<\infty$, and  $0 \leq \phi \leq 2 \pi$.	$M$ and $J$ are respectively the mass and angular momentum of the black hole. $L$ represents the radius of curvature of spacetime and satisfy $L^{2}=-1/\Lambda$ with $\Lambda$ being the negative cosmological constant. The metric is singular at the inner and outer horizons where $N^{2}(r)$ vanishes. Using gravitational path integral and saddle point approximation, the generalized off-shell free energy of BTZ black hole reads \cite{BTZ,York}
	\begin{equation}
		F=M-\frac{S}{\tau}+N^{\phi}(r_{h})J,\label{generallizedF}
	\end{equation}
	The parameter $\tau$ is an extra variable with the dimension of time. It varies freely and can be thought as the inverse temperature of the cavity enclosing the black hole.   $r_{h}$ represents the radius of the outer horizon. The mass and entropy can be written as the function of $r_{h}$ and $J$
	\begin{eqnarray}
		M&=&\frac{r_{h}^{2}}{L^2}+\frac{J^2}{4r_{h}^{2}},\label{M}\\
		S&=&4\pi r_{h}.\label{S}
	\end{eqnarray}
	We have to stress that $L$ is no more a fixed value so that the modified Bekenstein-Smarr mass formula still holds in 3-dimensional black hole thermodynamics \cite{newSmarr06,newSmarr07}. However, as we will show in this paper, its value will not affect the topological charge of BTZ solution. With the help of \eqref{M} and \eqref{S}, the generalized free energy \eqref{generallizedF} can be written as
	\begin{equation}
		F=\frac{r_{h}^{2}}{L^{2}}-\frac{J^2}{4r_{h}^{2}}-\frac{4\pi r_{h}}{\tau}.
		\label{F}
	\end{equation}
	It is off-shell except at $\tau=1/T$ which means the black hole is in the maximal mixed state.
	\section{Topological classes of BTZ black hole solutions}
	When the generalized free energy is obtained, we could apply the method in \cite{Liu22} to estabilish a parameter space and find the zero point of the vector field in it. Profoundly, the zero points are exactly corresponding to the on-shell black hole solution. We can calculate the topological number of them by virtue of Duan’s $\phi$-mapping topological current theory \cite{Duan}. The number can be seen as a characteristic value of the on-shell black hole solution. Following the spirit of \cite{Liu22}, we define the vector as
	\begin{equation}
		\phi=(\phi^{r_{h}},\phi^{\theta})=(\frac{\partial F}{\partial r_{h}},-\cot\theta \csc\theta)
	\end{equation}
	with $\theta\in[0,\pi]$ for convenience. The vector field is on $\theta-r_{h}$ space, and we can see that $\phi^{\theta}$ is divergent when $\theta=0,\pi$, making the direction of vectors point vertically outward at this boundary. The zero point, corresponding to $\tau=1/T$ \cite{Liu21}, can only be obtained when $\theta=\pi/2$.
	Now we introduce the topological current as
	\begin{equation}
		j^\mu=\frac{1}{2 \pi} \epsilon^{\mu \nu \rho} \epsilon_{a b} \partial_\nu n^a \partial_\rho n^b, \quad \mu, \nu, \rho=0,1,2
	\end{equation}
	where $\partial_\nu=\left(\partial / \partial x^\nu\right)$ and $x^\nu=\left(\tau, r_h, \theta\right)$. The unit vector is defined as $n^a=\left(\phi^a /\|\phi\|\right)(a=1,2)$. The conservation law of the current, $\partial_\mu j^\mu=0$ can be easily obtained by the definition of $j^\mu$, and $\tau$ here serves as a time parameter of the topological defect. By using the Jacobi tensor $\epsilon^{a b} J^\mu(\phi / x)=$ $\epsilon^{\mu \nu \rho} \partial_\nu \phi^a \partial_\rho \phi^b$ and the two-dimensional Laplacian Green function $\Delta_{\phi^a} \ln \|\phi\|=2 \pi \delta^2(\phi)$, the topological current can be written as
	\begin{equation}
		j^\mu=\delta^2(\phi) J^\mu\left(\frac{\phi}{x}\right) .
	\end{equation}
	where $j^\mu$ is nonzero only at $\phi^a\left(x^i\right)=0$, and we denote its $i$-th solution as $\vec{x}=\vec{z}_i$. The density of the topological current reads \cite{density}
	\begin{equation}
		j^0=\sum_{i=1}^N \beta_i \eta_i \delta^2\left(\vec{x}-\vec{z}_i\right),
	\end{equation}
	where $\beta_i$ is Hopf index, which counts the number of the loops that $\phi^a$ makes in the vector $\phi$ space when $x^\mu$ goes around the zero point $z_i$. Thus Hopf index is always positive. $\eta_i$ is the Brouwer degree and satisfy $\eta_i=$ $\operatorname{sign}\left(J^0(\phi / x)_{z_i}\right)=\pm 1$. Given a parameter region $\Sigma$, the corresponding topological number can be obtained as
	\begin{equation}
		W=\int_{\Sigma} j^0 d^2 x=\sum_{i=1}^N \beta_i \eta_i=\sum_{i=1}^N w_i,
		\label{W}
	\end{equation}
	where $w_i$ is the winding number for the $i$-th zero point of $\phi$ contained in $\Sigma$ and its value does not depend on the shape of the region where we perform the calculation. Usually, distinct zero points of the vector field are isolated, making Jacobian $J_0(\phi / x) \neq 0$. If Jacobian $J_0(\phi / x) = 0$, it means that the defect bifurcates \cite{Duan2}. Eq.\eqref{W} shows that in any given region, the global topological number is the sum of the winding number of each zero point which reflects the local property of the topological defect.
	
\begin{figure}
		\centering
		\includegraphics{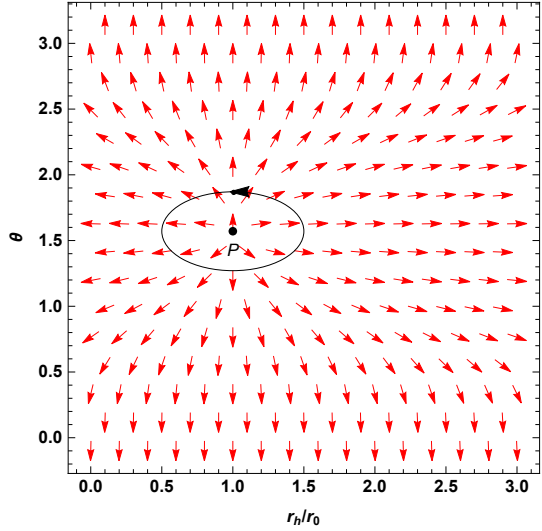}
		\caption{Unit vector field in parameter space with  $J=0$, $Q=0$ and $L=1/r_{0}$ solution as $\tau=2\pi/r_{0}^{3}$. $P$ marked with black dot at $(r_{h}/r_{0},\theta)=(1,\pi/2)$ is the zero point of the vector field.  The black contour is a closed loop enclosing the zero point and we can performing the calculation of Eq.\eqref{W} inside it. The shape and the size of the loop will not affect the winding number.
		}
		\label{fig1-1}
	\end{figure}

	Based on the approach introduced above, we first investigate BTZ black hole with $J=0$. In this case, the generalized free energy is
	\begin{equation}
		F=\frac{r_{h}^{2}}{L^2}-\frac{4\pi r_{h}}{\tau}.
	\end{equation}
	We define the vector field as
	\begin{equation}
		\phi=(\phi^{r_{h}},\phi^{\theta})=(\frac{2r_{h}}{L^2}-\frac{4\pi}{\tau},-\cot\theta \csc\theta).
	\end{equation}
	By solving the equation $\phi=0$, we acquire the relation
	\begin{equation}
		\tau=\frac{2\pi L^2}{r_{h}}, \quad \theta=\frac{\pi}{2}.
	\end{equation}
	We take $L=1/r_{0}$ and $\tau=2\pi/{r_{0}^{3}}$ with $r_{0}$ an arbitrary positive constant, so there is one zero point in $\theta-r_{h}$ plane at $(r_{h}/r_{0},\theta)=(1,\pi/2)$. We plot the unit vector field in Fig. \ref{fig1-1}. The loop surrounding the zero point sets the boundary of a given region so we can use Eq.\eqref{W} to acquire the winding number of point $P$. We find that $w=1$. Since there is only one defect in the parameter space, it gives rise the global topological number $W=1$ for BTZ black hole solution with $J=0$.

\begin{figure}
		\centering
		\includegraphics{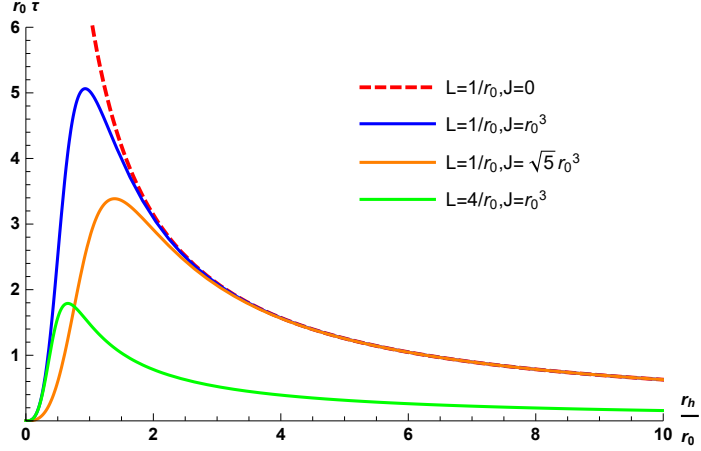}
		\caption{Solution curves in $\tau-r_{h}$ plane with different values of $L$ and $J$. There is always one turning point in the curve as long as $J\neq0$.
		}
		\label{fig2}
	\end{figure}

\begin{figure}
		\centering
		\includegraphics{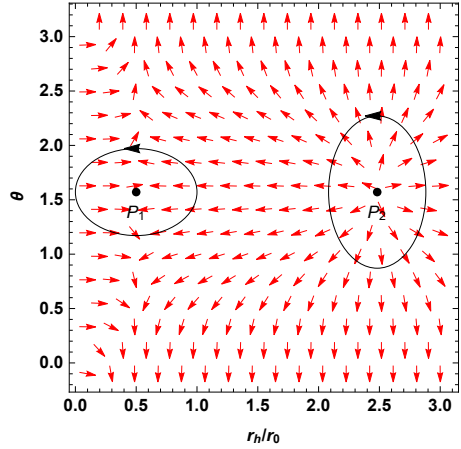}
		\caption{Vector field in $\theta-r_{h}$ plane with $J\neq0$. We take $\tau=4\pi/{5r_{0}^{3}}$. The points marked in black dot are zero points of the field. They are respectively at $P_{1}=(r_{h}/r_{0},\theta)=(0.5,\pi/2)$ and $P_{2}=(r_{h}/r_{0},\theta)=(2.482,\pi/2)$.
		}
		\label{fig2-1}
	\end{figure}

	Now we would further explore the rotating BTZ black hole. The generalized free energy is
	\begin{equation}
		F=\frac{r_{h}^{2}}{L^2}-\frac{4\pi r_{h}}{\tau}-\frac{J^2}{4r_{h}^{2}}.
	\end{equation}
	As a result, we define the vector field as
	\begin{equation}
		\phi=(\phi^{r_{h}},\phi^{\theta})=(\frac{2r_{h}}{L^2}-\frac{4\pi}{\tau}+\frac{J^2}{2r_{h}^{3}},-\cot\theta \csc\theta).
	\end{equation}
	Solving the equation $\phi=0$, we get the relation
	\begin{equation}
		\tau=\frac{8\pi r_{h}^{3}}{4L^{-2} r_{h}^{4}+J^{2}}, \quad \theta=\frac{\pi}{2}.
	\end{equation}
	Upon the value of $L$ and $J$ are determined, we could acquire a curve in $\tau-r_{h}$ plane. For instance, as shown in Fig. \ref{fig2}, when taking $L=1/r_{0}$ and $J=r_{0}^{3}$, the curve meets a peak at $(r_{h},\tau)\approx(0.93r_{0},5.06/r_{0}^{3})$ and rapidly goes down as $r_{h}$ gets larger. Consequently, the vector field possesses two zero points for small $\tau$, in comtrast to one zero point for non-rotating BTZ black hole case. As $\tau=4\pi/{5r_{0}^{3}}$, the two intersection points are respectively at $r_{h}=0.5r_{0}$ and $r_{h}\approx2.482r_{0}$. We illustrate the vector field and the zero points in Fig. \ref{fig2-1}. When $\tau=\tau_{cri}\approx5.06/r_{0}^{3}$, the intersection points coincide, and for larger $\tau$, annihilate. It is easy to check the critical point satisfy $d^{2}\tau/dr_{h}^{2}<0$, which belongs to annihilation point. For $\tau<\tau_{cri}$, we find that the winding number of the two zero points are $w_{1}=-1$ and $w_{2}=1$. Thus the global topological number for BTZ rotating solution is $W=w_{1}+w_{2}=0$, different from non-rotating case.

	When taking electric charge into account, the squared lapse $N^2(r)$ and the angular shift $N^\phi(r)$ are respectively given by
	\begin{eqnarray}
		N^2(r)&=&-M+\frac{r^2}{L^2}+\frac{1}{2}QA_{0}(r),\\
		N^\phi(r)&=&0
	\end{eqnarray}
	$Q$ denotes the electric charge of the black hole, $A_{0}(r)$ is the only nonvanishing component of the electromagnetic vector potential and is taken to be $A_{0}(r)=-Q \ln (r/r_{c})$ with $r_{c}$ being an arbitrary constant. the generalized off-shell free energy of BTZ black hole reads \cite{BTZ,York}
	\ba
		F&=&M-\frac{S}{\tau}-A_{0}(r_{h})Q\nn\\
&=&=\frac{r_{h}^{2}}{L^{2}}-\frac{4\pi r_{h}}{\tau}+\frac{1}{2}Q^{2}\ln\frac{r_{h}}{r_{c}}.\label{generallizedFQ}
	\ea
Following the same step, we get the on-shell solution curve in $\tau-r_{h}$ plane, which satisfy
	\begin{equation}
		\tau=\frac{8\pi r_{h}^{3}}{4L^{-2}r_{h}^{4}+Q^2r_{h}^{2}}.
	\end{equation}
	
	\begin{figure}
		\centering
		\includegraphics{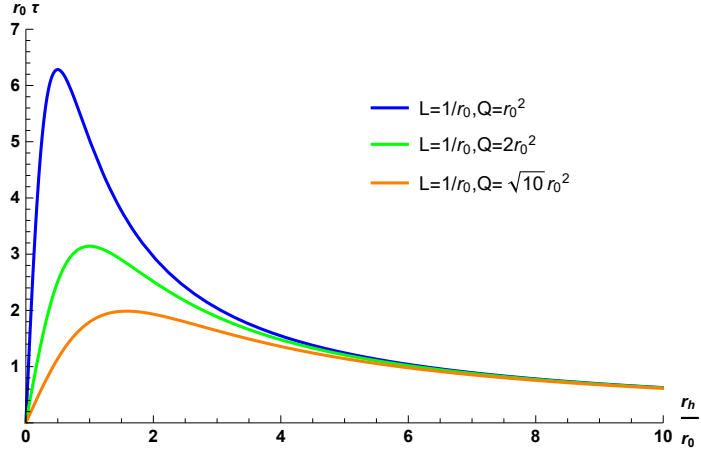}
		\caption{Solution curve in $\tau-r_{h}$ plane with different value of $L$ and $Q$. We find that non-trivial $Q$ do not bring essential difference to the trend of the curve compared with rotational case.
		}
		\label{fig3}
	\end{figure}
\begin{figure}
		\centering
		\includegraphics{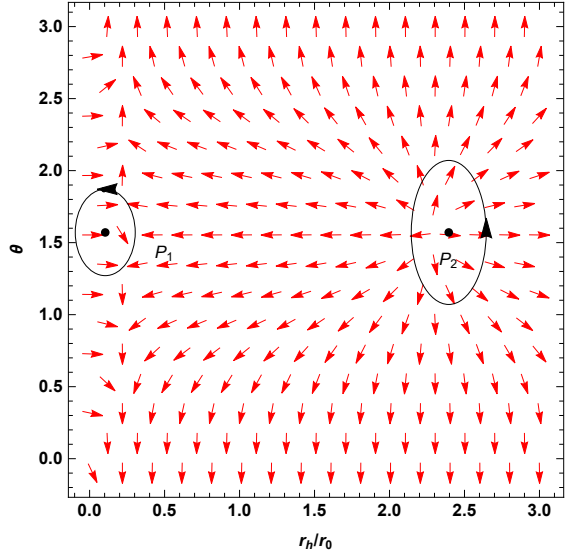}
		\caption{Vector field in $\theta-r_{h}$ plane with $Q\neq0$. We also take $\tau=4\pi/{5r_{0}^{3}}$. The zero points of the field are respectively at $P_{1}=(r_{h}/r_{0},\theta)=(0.104,\pi/2)$ and $P_{2}=(r_{h}/r_{0},\theta)=(2.395,\pi/2)$.
		}
		\label{fig3-1}
	\end{figure}

	Interestingly, we find that the electric charge do not change the rough trend of the curve. As shown in Fig. \ref{fig3}, there is invaribly one anihilation point in $\tau-r_{h}$ plane, and $\tau_{cri}$ merely gets smaller as the value of $Q$ taken larger. There are two zero points as well. We calculate the winding number of them for a given $\tau$ and find $w_{1}=-1$, $w_{2}=1$. See Fig. \ref{fig3-1} as an example. The global topological number of charged BTZ black hole solution is $W=0$, which is the same as the case of rotational BTZ black hole. Hence from the perspective of topological charge, the two kinds of 3-dimensional black hole are just the same. This conclusion is the same as four dimensional cases \cite{Kerr}. Yet the topological number of non-rotating and uncharged black hole are different, with $W_{BTZ}=1$ whereas $W_{Schwarzchild}=-1$. Besides, in contrast to the charged Reissner-Nordstrom anti de-Sitter (RN-AdS) black hole in four dimension \cite{Liu22}, three dimensional black hole solution evidently has fewer zero points. So the dimension of black hole of the same kind may have an unique influence on the quantity of defects and the global topological number.

	\section{Conclusion}
	In this paper, we use the generalized free energy of BTZ black hole to define a vector field in a parameter space $\theta-r_{h}$. We find the zero point of the field and obtain the winding number by applying Duan's $\phi$-mapping topological current theory. It is discovered that the unique black hole solution in three dimensional general relativity has one zero point for $Q=J=0$ and two for rotating or charged cases. The global topological charges are one and zero respectively. We also find that the dimension of the AdS background would lead to a distinct amount of zero points.

The previous works \cite{Liu22,Kerr} indicate that all black hole solutions in the pure Einstein-Maxwell gravity theory should be classified into three different topological classes for four and higher spacetime dimensions. This observation is further enhanced in the modified gravity case \cite{GB,Lovelock}. However, our investigation on BTZ black holes found only two topological classes for BTZ spacetime. This means this feature is not universal and the spacetime dimension seems strongly relevant in the topological classification of black holes.
	
	There are many issues that deserve further investigation. Generalize our results to Kerr-AdS and Kerr-dS and compare with the existing result will be interesting. Moreover, another interesting object is to investigate the topological number of the black hole solutions in the supergravity and modified gravity theories. We leave these interesting topics for future studies.

\begin{acknowledgements}
We would like to thank Prof. Pujian Mao for helpful discussion. This work is supported by NSFC with Grants No.12275087 and ``the Fundamental Research Funds for the Central Universities''.
\end{acknowledgements}	
	\appendix
	\par
	\bibliographystyle{unsrt}

\end{document}